\documentstyle[12pt]{article}
\begin{document}
\begin{center}{\large\bf QUALITATIVE CONSIDERATION OF THE EFFECT OF
BINDING IN DEEP INELASTIC SCATTERING}\end{center}
\vskip 1em \begin{center} {\large Felix M. Lev} \end{center}
\vskip 1em \begin{center} {\it Laboratory of Nuclear
Problems, Joint Institute for Nuclear Research, Dubna, Moscow region
141980 Russia (E-mail:  lev@nusun.jinr.dubna.su)} \end{center}
\vskip 1em
\begin{abstract}
 In our recent paper (hep-ph 9501348) we argued that the Bjorken
variable $x$ in deep inelastic scattering cannot be interpreted as the
light cone momentum fraction $\xi$ even in the Bjorken limit and in zero
order of the perturbation theory. The purpose of the present paper is to
qualitatively explain this fact using only a few simplest kinematical
relations.
\end{abstract}
\vskip 1em

 Let us consider the deep inelastic electron (or muon) scattering off a
nucleon. Let $P'$ be the 4-momentum of the initial nucleon and $q$ be
the 4-momentum of the virtual photon absorbed by this nucleon. The
Bjorken limit of deep inelastic scattering (DIS) is the case when $|q^2|$
and $(P'q)$ are very large but $x=|q^2|/2(P'q)$ is not too close to 0
or 1.

 In the framework of the Feynman parton model [1] the Bjorken variable
$x$ has the following simple interpretation. Consider the infinite
momentum frame (IMF) in which the nucleon moves along the positive
direction of the $z$ axis. Suppose that the nucleon in the IMF can be
considered as a system of almost free partons with the 4-momenta $p_i'$
$(i=1,2...)$. Define the quantity $\xi_i=(p_i^{'0}+p_i^{'z})/
(P_i^{'0}+P_i^{'z})$. Then, if the virtual photon is absorbed by the
$i$-th parton, the quantity $\xi_i$ is equal to $x$ in the Bjorken
limit. This fact is explained in many textbooks and papers, and it is
shown that $\xi_i=x$ only in zero order in $\alpha_s$ (where $\alpha_s$
is the QCD running coupling constant) since the perturbative QCD
corrections leads to the logarithmic breaking of this relation.

 The question arises whether the perturbative consideration of the
interparton interactions is compatible with the fact that the partons
form the bound state --- the nucleon --- which cannot be considered in the
framework of the perturbation theory. A very exact characteristic of
this situation is given in the following extract from Ref. [2]: "The
parton model replaces unknown theory of hadronic bound states, while
such a theory should in principle follow from QCD. Perturbation theory
based on asymptotic freedom --- the basis of success of QCD --- does not
apply in the case of bound state problems and the theory is in a corner."

 In Ref. [3] the effect of binding in DIS was considered using the
exact solution for the electromagnetic current operator (ECO) found in
Ref. [4], and it was shown that $\xi_i\neq x$ even in the Bjorken limit
and in zero order of the perturbation theory. Then the interpretation
of the DIS data considerably differs from the usual interpretation
(see Ref. [3] for details). Probably for this reason the opinion of
several physicists is that the results of Ref. [3] are wrong, or at
best, the found solution is correct but it is a pathology (in our
opinion, even in this case the solution is of interest). Taking into
account this criticism, we find it useful to give a short qualitative
consideration of our results.

 Let us consider a system of $N$ free particles with the masses $m_i$
and the 4-momenta $p_i$ $(i=1,...N)$. The number $N$ can be arbitrary
including $N=\infty$. The 4-momenta $p_i$ have the components
$(\omega_i({\bf p}_i),{\bf p}_i)$ where $\omega_i({\bf p}_i)=
(m_i^2+{\bf p}_i^2)^{1/2}$. Each 4-momentum $p_i$ is fully determined
by the ordinary momentum ${\bf p}_i$.

 Instead of the individual variables ${\bf p}_i$ we can introduce the
total momentum ${\bf P}={\bf p}_1+...{\bf p}_N$, while the internal
momentum variables ${\bf k}_i$ $(i=1,...N)$ can be defined as follows.
First we introduce the free mass operator $M_0$ as
$M_0=[(p_1+...p_N)^2]^{1/2}$. Then we define the Lorentz boost
$L({\bf P}/M_0)$ from the c.m. frame of the system under consideration
to the reference frame where the total momentum of this system is equal
to ${\bf P}$. The explicit form of $L({\bf P}/M_0)$ is not important
for our consideration, but it is important that the boost is fully
determined by the vector ${\bf P}/M_0$. Finally we define the 4-vectors
$k_i=(\omega_i({\bf k}_i),{\bf k}_i)$ as
\begin{equation}
k_i=L(\frac{{\bf P}}{M_0})^{-1}p_i
\end{equation}
It is easy to show that ${\bf k}_1+...{\bf k}_N=0$ as it should be, and
therefore only $N-1$ vectors ${\bf k}_i$ are independent.

 Conversely, if the vectors ${\bf k}_1,...{\bf k}_N$ and ${\bf P}$ are
known, we can define the mass operator $M_0$ as $\omega_1({\bf k}_1)+
...\omega_N({\bf k}_N)$ and then the $p_i$ are given by
\begin{equation}
p_i=L(\frac{{\bf P}}{M_0})k_i
\end{equation}

 The Hilbert space $H$ of states for the system under consideration
is the space of functions
$\varphi({\bf p}_1,...{\bf p}_N,\mbox{spin variables})$ quadratically
integrated over some measure. We can also introduce the internal
Hilbert space $H_{int}$ as the space of functions
$\chi({\bf k}_1,...{\bf k}_N,\mbox{spin variables})$ and represent
$H$ as the space of functions $\Phi({\bf P})$ with the range in $H_{int}$.
It is obvious that this construction can be done not only if $N$ is
fixed but also in the case of quantum field theory when the elements
of the Hilbert spaces are some Fock columns.

 Let us now consider the case when the particles interact with each
other. Then the mass operator ${\hat M}$ is the operator in $H_{int}$.
If the system is in the bound state with the mass $M'$, its internal
wave function $\chi$ is the eigenfunction of ${\hat M}$ with the
eigenvalue $M'$: ${\hat M}\chi=M'\chi$.

 It is obvious that the quantities $p_i$ no longer can be interpreted
as the 4-momenta of the corresponding particles if they interact with
each other. Of course, in the presence of the interaction the Hilbert
space $H$ remains the same as for noninteracting particles, i.e. we can
still use the realization of $H$ as the tensor product of the
single-particle states, but it is reasonable to expect that the
collective variables ${\bf k}_1,...{\bf k}_N$ and ${\bf P}$ are more
convenient than ${\bf p}_1,...{\bf p}_N$.

 If ${\bf P}=0$ then, as follows from Eq. (2), $p_i=k_i$. For this
reason one might think that $k_i$ still has the meaning of the
4-momentum of particle $i$ in the c.m. frame. If the system is in the
bound state with the mass $M'$, we can define the 4-momenta
\begin{equation}
h_i=L(\frac{{\bf P}}{M'})k_i
\end{equation}
Since $M'\neq M_0$, it is obvious from Eqs. (2) and (3) that
$h_i\neq p_i$. The boost entering into Eq. (3) is the real physical
boost since ${\bf P}$ and $M'$ are the real momentum and mass of the
system as a whole. At the same time, the boost entering into Eq. (2)
is now unphysical. For this reason one might think that $h_i$ has the
meaning of the 4-momentum of particle $i$ in the reference frame where
the total momentum of the system is equal to ${\bf P}$.

 It is reasonable to think that $h_i$ is a more appropriate candidate
for the role of the 4-momentum of particle $i$ than $p_i$, but strictly
speaking none of them can be interpreted in such a way in the presence
of the interaction.

 Now we return to DIS. The usual words about this process are that the
absorption of the virtual photon with the large momentum is so quick,
and the internal hadronic clocks in the IMF are so slow that if the
photon is absorbed by parton $i$, the states of the spectator partons do
not change in the process of absorption. But how can we define these
states? For example, we can assume that if $p_l'$ are the parton
momenta in the initial system, and $p_l"$ are the same momenta in the
final system $(l=1,...i-1,i+1,...N)$ then
\begin{equation}
P'+q=P",\quad p_l'=p_l"
\end{equation}
where $P"$ is the total 4-momentum of the final system. The first
expression in Eq. (4) is the total 4-momentum conservation (thus it
should be satisfied in any case), but the second one needs
substantiation. In view of the above discussion one might think that
the conditions
\begin{equation}
P'+q=P",\quad h_l'=h_l",
\end{equation}
where $h_l'$ and $h_l"$ are the quantities $h_l$ in the initial and
final system, are not less reasonable than the conditions (4).

 It is easy to show (see Refs. [1,3] and references quoted therein) that
Eq. (4) leads to the well-known result $\xi_i=x$ in the Bjorken limit,
while Eq. (5) leads to the relation between $\xi_i$ and $x$ derived in
Ref. [3].

 In principle, the relations (4) or (5) should not be imposed "by hands",
but they should automatically follow from the form of the ECO. If the
ECO is taken in the impulse approximation (IA), i.e. as a sum of the
constituent ECO's, the immediate consequence of such a choice is
obviously Eq. (4). However, as pointed out by many authors, the ECO in
the IA does not satisfy even relativistic invariance (see, for example,
Ref. [5]). On the
other hand, Eq. (5) is the consequence of such a choice for the ECO
when it satisfies relativistic invariance and current conservation [3].
Therefore Eq. (5) is indeed more reasonable than Eq. (4).

 Let us stress once more that the only dynamics involved in Eqs. (4) and
(5) is that the initial bound state has the mass $M'\neq M_0$; in
particular these equations
do not depend on whether the number of particles is finite or
infinite and whether the asymptotic freedom takes place. Equation (5)
was obtained in Ref. [3] simply because for the ECO satisfying
relativistic invariance and current conservation the mass entering
into the Lorentz boost is automatically equal to the real physical
mass while the choice of the IA automatically leads to the fact that
the corresponding Lorentz boost depends on $M_0$ [3]. For this reason
we believe that though the choice of the ECO satisfying relativistic
invariance and current conservation is not unique [4], and the
expression for the hadronic tensor derived in Ref. [3] is
model-dependent, Eq. (5) is model-independent if the states of the
spectator partons do not change in the process of absorption of the
virtual photon.

 The reader can say that it is difficult to believe that the relation
$\xi_i=x$ in the Bjorken limit and in zero order of the perturbation
theory may be invalid since it was derived by many authors and in
different approaches. However in all these approaches the ECO
in zero order of the perturbation theory is the IA . Such an ECO
corresponds to the case when the constituents comprising the nucleon are
free, while zero order of the perturbation theory should be compatible
with the fact that the nucleon is the bound system.

 Our experience in conventional nuclear and atomic physics tells that
the IA is a good approximation at large momentum transfer, but the
corresponding calculations agree with the data only in the
nonrelativistic approximation. In this approximation there is no
difference between our approach and the IA since $M_0$ and $M'$ are
equal to each other in the nonrelativistic case.

 In conclusion we briefly discuss the following question. It is
well-known that the perturbation theory does not apply to the bound
state problem. However, it is believed that the electromagnetic
processes involving relativistic bound states can be reliably
calculated using only a few Feynman diagrams. Meanwhile, if we expand
the ECO in powers of $\alpha_s$, the same should be done with the
bound states, but this is not justified. Our solution for the ECO
which leads to Eq. (5) automatically implies that there do not exist
any finite sets of the Feynman diagrams which describe the
electromagnetic processes involving relativistic bound states with a
good accuracy. Indeed, this solution shows that a rather simple
description of such processes can be obtained in the variables
$h_i$, while the Feynman diagrams describe the processes in the
variables $p_i$.

\vskip 1em
\begin{center} {\bf Acknowledgments} \end{center}

\begin{sloppypar}
 The author is grateful to I.L.Grach, I.M.Narodetskii, Y.N.Uzikov  and
H.J.Weber for valuable discussions, and to S.J.Brodsky, F.Coester,
L.Frankfurt, S.D.Glazek, B.L.Ioffe, L.A.Kondratyuk and M.P.Locher for
useful remarks. This work was supported by grant No. 93-02-3754 from the
Russian Foundation for Fundamental Research.
\end{sloppypar}


\begin{thebibliography}{20}
\bibitem{Feyn} R.P.Feynman, Phys.Rev.Lett. {\bf 23}, 1415 (1969);
J.D.Bjorken and E.A.Paschos, Phys.Rev. {\bf 185}, 1975 (1969).
\bibitem{Glazek} S.D.Glazek. "Relativistic Bound States of Elementary
Particles in Light-Front Hamiltonian Approach to Quantum Field Theory".
Preprint IFT/7/93, Warsaw (1993).
\bibitem{l1} F.M.Lev, hep-ph 9501348 (1995).
\bibitem{lev} F.M.Lev, Annals of Physics (N.Y.) {\bf 237}, 355 (1995).
\bibitem{eco} J.L.Friar, Annals of Physics (N.Y.) {\bf 104}, 380 (1977),
Phys.Rev. {\bf C22}, 796 (1980); W.Bentz, Nucl.Phys. {\bf A446}, 678
(1985); F.Gross and D.O.Riska, Phys.Rev. {\bf C36}, 1928 (1987);
A.Yu.Korchin and A.V.Shebeko, Yad.Fiz. {\bf 54}, 357 (1991); F.Coester,
Progr.Part.Nucl.Phys. {\bf 29}, 1 (1992).
\end{thebibliography}
\end{document}